\documentclass[11pt]{article}
\usepackage{moriond,epsfig}

\def\be{\begin{equation}}
\def\ee{\end{equation}}
\def\bea{\begin{eqnarray}}
\def\eea{\end{eqnarray}}

\begin{document}
\vspace*{4cm}
\title{Multiparameter estimation with  the Pseudo-$C_l$
method\footnote{Presented by K.~M.~G\'orski}}

\author{ Benjamin D.\ Wandelt$^{1,2}$, Krzysztof M.\
G\'orski$^{1,3,4}$ and  Eric Hivon$^{1,5}$ 
}

\address{$^1$Theoretical Astrophysics Center,
        Juliane Maries Vej 31,
        DK-2100 Copenhagen \O,
        Denmark\\
        $^2$Department of Physics,
        Princeton University,
        Princeton, NJ 08540,
        USA\\
	$^3$European Southern Observatory, Garching bei M\"unchen, Germany\\
        $^4$Warsaw University Observatory, Warsaw, Poland\\
	$^5$Observational Cosmology,
	Caltech,  Mail Code 59-33,
	Pasadena, CA 91125,
	USA
        }

\maketitle\abstracts{We apply the Pseudo-$C_l$ formalism to obtain an
unbiased, approximate method for 
efficient
simultaneous estimation of several cosmological parameters from large,
almost full--sky cosmic
microwave background data sets.}

\section{Introduction}

Within the standard model of cosmology there are about 10 parameters
which characterise the properties of our
Universe. It is one of the key goals of future CMB experiments such as MAP and
Planck to determine these cosmological parameters to high precision. This undertaking faces the challenge that
realistic CMB data is necessarily incomplete and noisy. The Galaxy
obscures roughly a third of the sky and because of the smallness of
the anisotropy signal, detector noise is not
negligible in the analysis. This leads to the computational challenge
which was expertly described at this meeting  in the contribution by
Borrill\cite{borrillMoriond}. 

In this talk we apply the pseudo-$C_l$ formalism\cite{WandeltHivonGorski} to this problem and show that it can be used
to develop an approximate form of the likelihood which has several useful
properties:
 it is Gaussian and hence easy to apply; it does not suffer from the usual disadvantages of Gaussian
approximations such as obtaining negative estimates of positive
definite quantities; it is computationally efficient with memory usage
of $N_{pix}$ 
and number of operations scaling as $N_{pix}^{\frac32}$ per likelihood
evaluation with a very small pre--factor leading to thousands of
likelihood evaluations per CPU hour.

\section{The approximation scheme}

Given a true CMB sky T and an experimental setup and observation strategy  (encoded in
the beam pattern B, the survey geometry $W$ and the noise distribution on the
sky $W_N T_N$) we can represent the observed temperature 
anisotropy map as
\begin{equation}
\tilde{T}({\bf \gamma})=W({{\bf {\gamma}}})\left[B\ast T({{\bf {\gamma}}})+W_N({{\bf {\gamma}}})T_N({{\bf {\gamma}}})\right]
\label{realtemp}
\end{equation}

This temperature field $\tilde{T}$ can be decomposed into spherical
harmonics coefficients 
\begin{equation}
\tilde{a}_{lm}  =  \int_{{\cal O}} d\Omega\: Y^{\ast}_{lm}({{\bf {\gamma}}}) \tilde{T}({{\bf {\gamma}}}).
\label{eq:palms}
\end{equation}
The notation ``$\int_{\cal O}$'' denotes integration over the fraction
of the sky covered by the survey.
These combine into the observed 
power spectrum coefficients which we call pseudo-$C_l$,
\begin{equation}
\tilde{C}_l=\frac1{2l+1}\sum_m
\left\vert{\tilde{a}_{lm}}\right\vert ^2.
\label{pcldef}
\end{equation}

In \cite{WandeltHivonGorski} we derive the exact statistics of the
pseudo-$C_l$, under the assumptions of
azimuthal survey geometry and noise which is uncorrelated from pixel to
 pixel and whose amplitude varies only from latitude to latitude.
The results we derived were still a superb approximation for strongly
non-azimuthal noise patterns. 

We found that in the case of large sky coverage the Pseudo-$C_l$
distributions were nearly indistinguishable from Gaussian distributions
of the same means and variances as long as $l>100$. The fact that many of the
cosmological parameters are sensitive to the power spectrum at
precisely these small scales led us to propose the following approximation
to the likelihood:

\begin{equation}
\widehat{\cal L}(C_l)=\prod_{l>100} \exp\left[-\frac{\left(C_l-\left\langle {\tilde{C_l}} \right\rangle\right)^2}{\left\langle {\Delta \tilde{C}_l^2} \right\rangle}\right]
\label{likelihood_approx_approx}
\end{equation}
In this approximation, maximum likelihood estimation has  reduced to simple $\chi^2$
fitting, however with the correct means and variances.

\begin{figure}[tf]
\centerline{\psfig{file=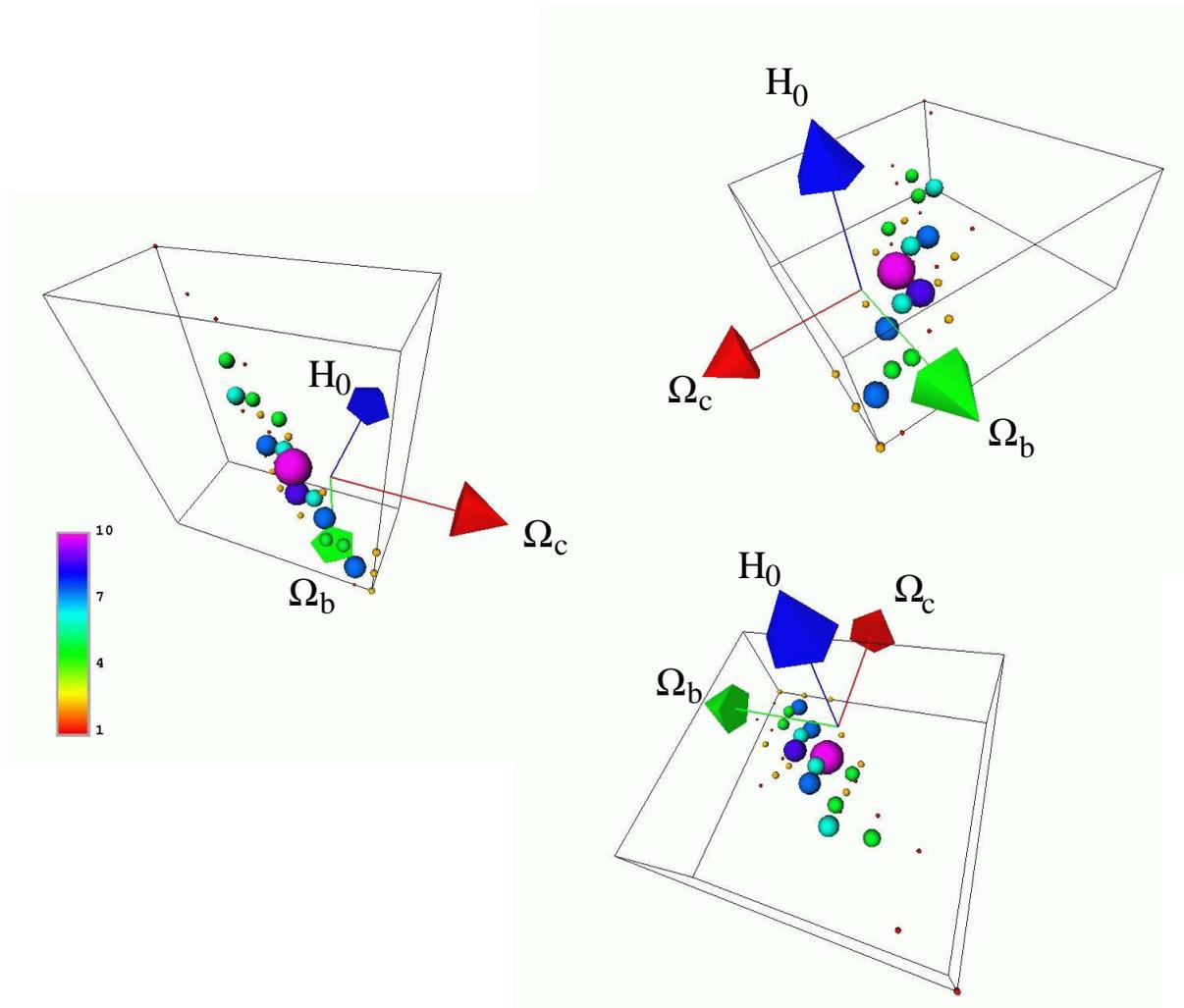,width=\textwidth}}
\caption[Multi--parameter estimation of $\Omega_0$, $\Omega_b$ and $H_0$ using  naive $\chi^2$ fitting]
{Multi--parameter estimation of $\Omega_0$, $\Omega_b$ and $H_0$ using
naive $\chi^2$ fitting. The three panels show 
three views from different directions of the
empirical distribution of the parameter estimates. 
Each sphere represents one bin of the
three--dimensional distribution. The size and shading of a sphere
indicates the number of realisations (out of 100 total) which led to parameter estimates
within its bin. The true
parameter values are at the origin of the coordinate axes. It is clearly
visible that the distribution is shifted with respect to the true
distribution by an amount which is inconsistent with the width of the
distribution. In other words the true values could be ruled out at
high significance if
this estimate was used.}
\label{stupid3D}
\end{figure}

\begin{figure}[tf]
\centerline{\psfig{file=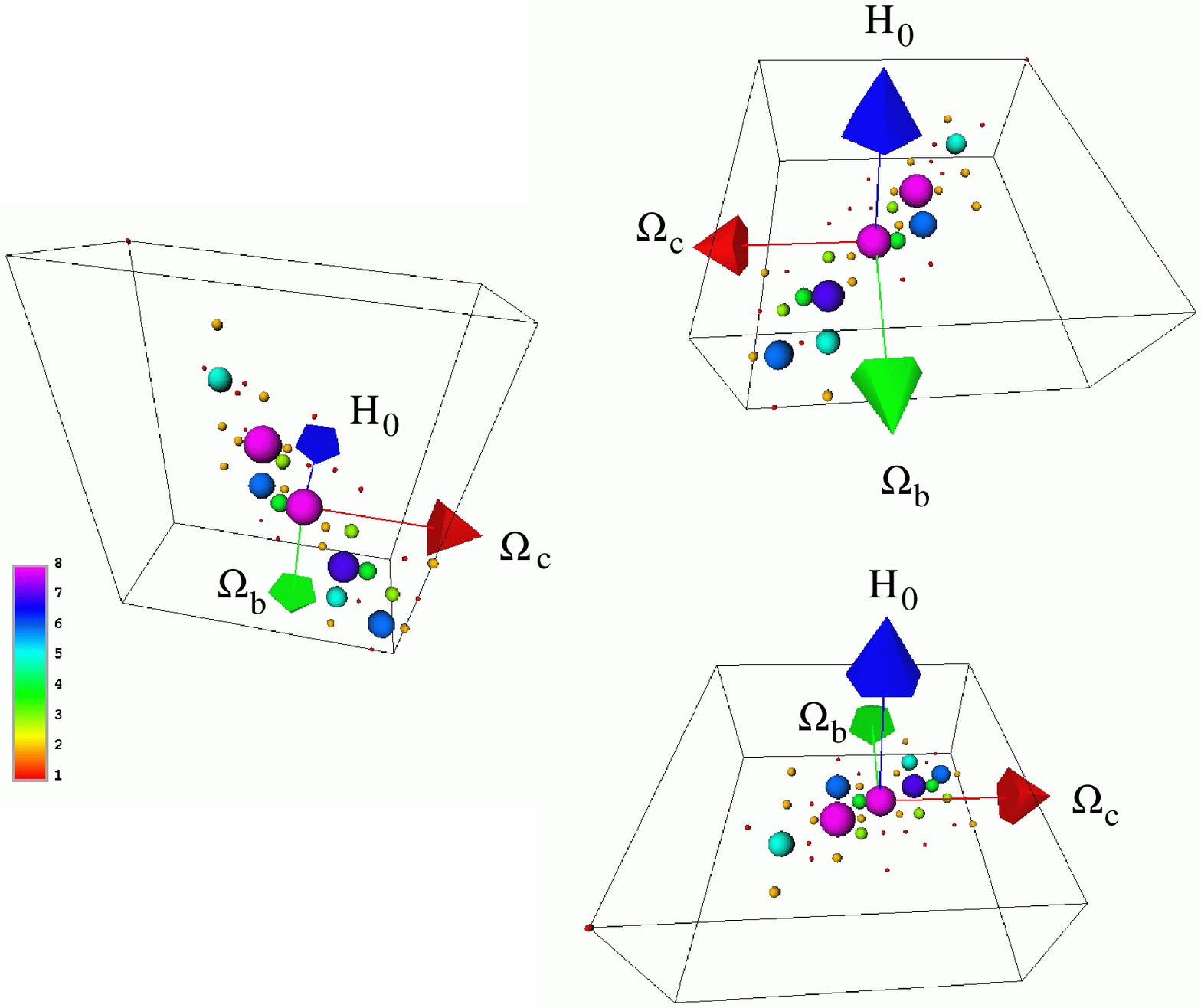,width=\textwidth}}
\caption[Multi--parameter estimation of $\Omega_0$, $\Omega_b$ and $H_0$ using our approximate likelihood.]
{Multi--parameter estimation of $\Omega_0$, $\Omega_b$ and $H_0$ using our approximate likelihood
 Eq.\ (\ref{likelihood_approx_approx}).  The three panels show
three views from different directions of the
empirical distribution of the parameter estimates. 
Each sphere represents one bin of the
three--dimensional distribution. The size and shading of a sphere
indicates the number of realisations (out of 100 total) which led to parameter estimates
within its bin. The true
parameter values are at the origin of the coordinate axes.  It is clearly
visible that the distribution is correctly centered on the true
values. }
\label{pcl3D}
\end{figure}

To illustrate, we solve the problem of estimating 3 parameters
($\Omega_c$ $\Omega_b$ and $H_0$) 
simultaneously from a sky with $12\times 10^6$ pixels of which $66\%$
are observed. The response of the experiment is modelled as a Gaussian
beam of FWHM 12 arcminutes. The  noise template is inhomogeneous and
not azimuthally symmetric with rms amplitude of $124\mu K$ per 3.5
arcminute pixel.
To compare with the naive Gaussian
approach and to show that our method is unbiased, we compute maximum (approximate) 
likelihood estimates from 100
realisations of the sky and plot a representation of the 
empirical distribution of parameter 
estimates in three dimensions in Figures \ref{stupid3D} (naive $\chi^2$)
and \ref{pcl3D} (our approach).

Our estimates are unbiased. The distributions of the
estimates are clearly centered on the true values. 

We stress that our approach avoids the usual difficulties of Gaussian approximations. For example, even  
though we use the Gaussian pdf, which  of course does not exclude
negative $C_l$,   
they are assigned an exceedingly small probability. This is because no
attempt is made to subtract out 
the noise contribution from the pseudo--$C_l$ --- instead it is
modelled consistently and the (signal$\times$noise) cross term which is present
in each realisation is not allowed to dominate.

\section*{References}

\begin{thebibliography}{99}
\bibitem{borrillMoriond} J.~Borrill 2000, in these proceedings 

\bibitem{WandeltHivonGorski}
B.~D.~Wandelt, E.~F.~Hivon, and K.~M.~G{\'o}rski.
\newblock 2000,
\newblock submitted to Physical Review D.
\end{thebibliography}

\end{document}